\pdfoutput=1
\documentclass[conference]{IEEEtran}
\usepackage{xcolor}
\usepackage{subfig}
\usepackage{graphicx}
\usepackage{siunitx}
\usepackage{lipsum}
\usepackage{comment}
\usepackage{floatrow}
\usepackage{amsfonts} 
\usepackage[utf8]{inputenc}
\usepackage[T1]{fontenc}
\usepackage{textcomp}

\begin{document}
\title{\bf Exploring the Role of Machine Learning\\ in Scientific Workflows:\\ Opportunities and Challenges}
\author{\IEEEauthorblockN{Azita Nouri\IEEEauthorrefmark{1},
Philip E. Davis\IEEEauthorrefmark{2},
Pradeep Subedi\IEEEauthorrefmark{2}, 
Manish Parashar\IEEEauthorrefmark{2}}
\IEEEauthorblockA{\IEEEauthorrefmark{1} \textit{School of Computer Science, Rutgers University, NJ, USA}\\
azita.nouri@rutgers.edu
}
\IEEEauthorblockA{\IEEEauthorrefmark{2} \textit{Scientific Computing Imaging Institute, University of Utah, Salt Lake City, UT, USA
}\\ 
philip.davis@sci.utah.edu,
pradeep.subedi@utah.edu,
parashar@sci.utah.edu
}}


\maketitle

\thispagestyle{empty}
\pagestyle{empty}

\begin{abstract}
In this survey, we discuss the challenges of executing scientific workflows as well as existing Machine Learning (ML) techniques to alleviate those challenges. We provide the context and motivation for applying ML to each step of the execution of these workflows. Furthermore, we provide recommendations on how to extend ML techniques to unresolved challenges in the execution of scientific workflows. Moreover, we discuss the possibility of using ML techniques for in-situ operations. We explore the challenges of in-situ workflows and provide suggestions for improving the performance of their execution using ML techniques.

\end{abstract}
\section{\textbf{Introduction}}
The past decade has seen a tremendous growth in Machine learning (ML) research as well as its use in a range of application domains. ML techniques can capture the patterns, relations, and behavior of different environments and datasets \cite{chen2014big}. They enable machines to automatically learn from data, identify patterns, build analytical models, and make decisions with minimal human intervention. ML processes involve a training phase through which the characteristics of the data as well as patterns in the data can be learned and used to make predictions for current or future datasets and events. ML algorithms combined with recent advances in high performance computing (HPC) have the potential for providing new and timely insights in a wide range of domains. Recent years have also seen an increasing interest in the use of ML as part of scientific workflows to accelerate scientific discovery ~\cite{baker2019workshop}. The growing volumes of data produced by experiments, observations as well as large-scale simulations, coupled with increasing computational power provide tremendous opportunities for leveraging ML techniques to improve and/or accelerate scientific insights~\cite{correa2018accelerating}. Scientific workflow use ML techniques in multiple ways. For example, ML techniques are used to extract the underlying structure \cite{nouri2021scalable}, construct predictive models, and help deal with the inherent complexity of the data \cite{nouri2021scalable}. ML techniques are also used to improve performance, resource provisioning, scalability, and reliability and robustness during workflow execution~\cite{nemirovsky2017machine}. 

The goal of this paper is to survey the use of ML techniques as part of scientific workflows running on large-scale systems. In this paper, we explore the landscape of ML techniques in the context of scientific workflows across a range of domains and usecases. We focus on different aspects of scientific workflows, including preparing the inputs, executing the workflow and processing its outputs, and study which ML techniques are used and how they are used in a workflow. We also explore associated challenges and opportunities. For example, in preparing inputs for a workflow, we explore the role of ML for input handling, parameter tuning, model optimization, computational steering, and workflow composition. During workflow execution, we explore the application of ML techniques for scheduling, failure detection and fault management, data placement and movement, I/O and memory management, and overall performance optimization. In handling the outputs from a workflow, we explore efforts that use ML techniques for correctness and verifiability of the results. We finally focus particularly on the use of ML techniques by in-situ scientific workflows, which become increasingly important at extreme scales.

Note that there have been other surveys of scientific workflows over the years that have primarily focused on identifying current and future challenges and providing possible solutions \cite{deelman2019role,Peterka2019}. This paper complements these surveys and focuses on the state of the art in the application of ML in scientific workflows. 

\begin{figure*}
\centering
    \includegraphics[width=0.9\textwidth,height=9cm]{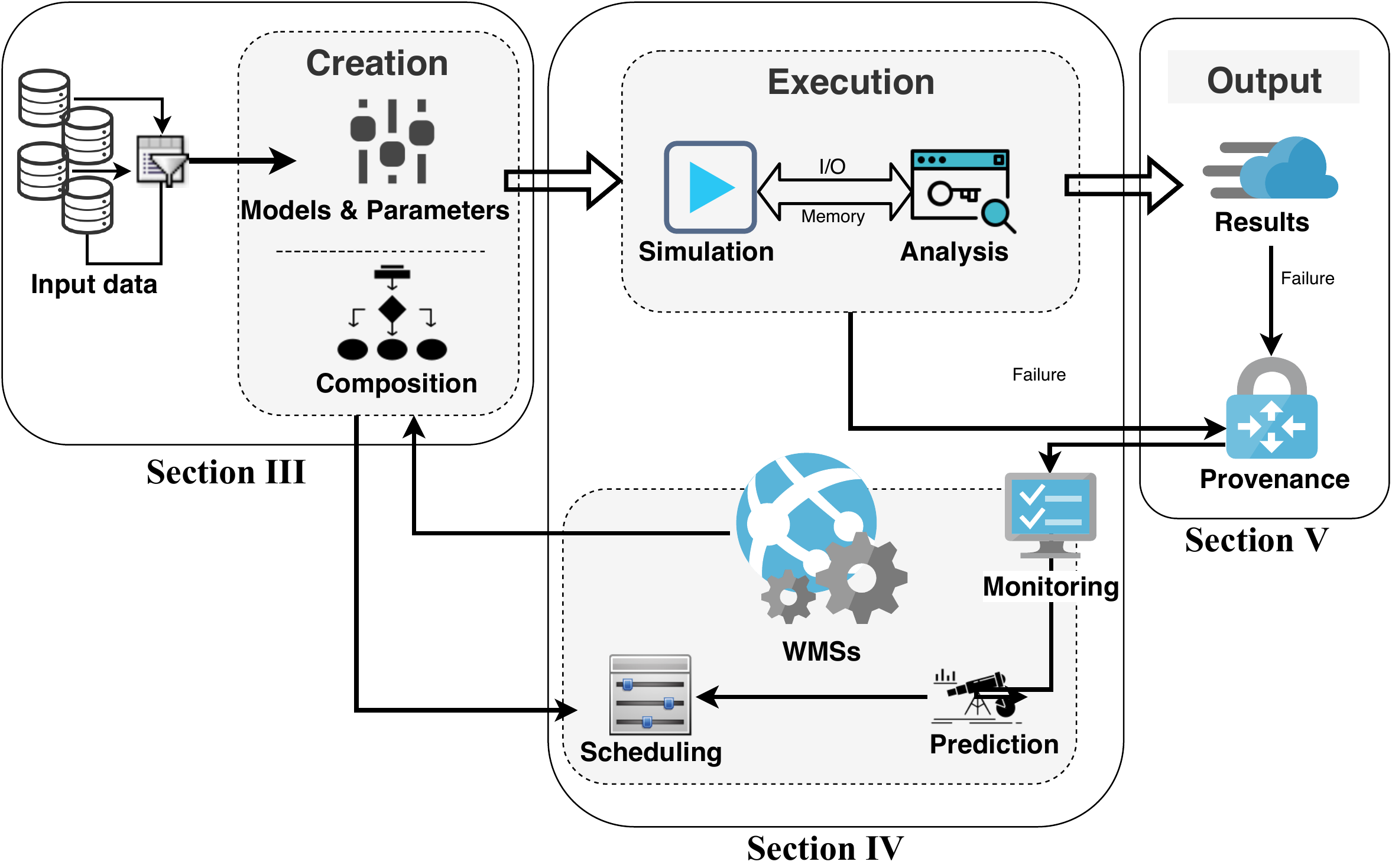}
    \caption{\textbf{Overview of the survey.} Illustration of how the workflow is composed of input, then executed and lastly how its output managed. Section number describe the following contents in order of appearance. For executing a scientific workflow, first data and model are prepared based on requirement of execution components. Then, simulation, and analysis/visualization form the core of workflow execution while WMSs perform vital operations, such as scheduling, monitoring, and resource allocation (I/O and memory) with the aim of optimizing the performance of workflow execution. And in the last part, how the output of the workflow is managed through keeping provenance information and/or interacting with WMSs.}
\label{fig:Cycle}
\end{figure*}

\subsection{Outline of the survey}
The following section provides the background context for this survey. To facilitate reading, a framework for the survey is presented in Figure \ref{fig:Cycle}. As illustrated in Fig.\ref{fig:Cycle}, we focus on three key aspects of a scientific workflow: creation, execution, and output processing. We start with input preparation, parameter- and model tuning, and workflow composition in Section III. As is shown in Fig.\ref{fig:Cycle}, workflow composition is typically a function provided by Workflow Managements Systems (WMSs) that enables users to modify a workflow even while it is executing. Section IV focuses on key runtime aspects of WMSs including scheduling, monitoring, prediction of resource usage, I/O and memory, and failure detection. Management of workflow outputs and provenance is discussed in Section V. Finally, in Section VI, we focus on in-situ workflows. We present the key challenges associated with the in-situ execution of scientific workflows and how current ML techniques can be used to address them. Section VII presents a summary and concludes the survey.
\section{\textbf{Background}}
\subsection{Scientific Workflows}
Scientific workflows have been used in many different domains including earth science \cite{turuncoglu2013toward}, astronomy \cite{habib2016hacc}, physics \cite{dolgert2008provenance}, and bioinformatics \cite{mehta2011enabling} where datasets are large and heterogeneous (from several sources and different formats). Scientific workflows represent the data flow and their dependencies in an acyclic graph, which typically represents complex computations and analytic. They include thousands of steps that involve diverse data flows and dependencies as well as different execution mechanisms in a distributed environment. Workflow management systems automate the execution of computational tasks in the next steps and define the order of these steps based on data flow and data dependency \cite{deelman-fgcs-2015}. The scientific workflows give high-level abstraction to the graph of dataflows between various components and simplify the automation and management of the workflow life cycle. In addition to automation, reproducibility, result sharing, and result derivation are necessary for collaborative researchers to have faster analysis and processes \cite{stodden2016enhancing}. Putting together, managing and representing such complex distributed systems have many challenges from creation, reproducibility, provenance, monitoring, performance optimization, reliability, robustness, scalability and intelligent support for workflow design \cite{Baker2018}.

\subsection{Machine Learning and Scientific Workflows}
Considering the growth in computing power, decreasing price of computational resources, as well as the larger scale of available data in different scientific domains, new techniques need to be provided to solve the challenges of scientific applications. ML is a good candidate as it is able to tackle complex problems (e.g. scheduling) and find interesting structures in large datasets relatively inexpensive. With the power of different learning algorithms, it is now possible to automatically model unstructured data, analyze larger scale, and more complex data \cite{chen2015supervised}. As a consequence, ML can filter useful information that leads to major advancements \cite{wozniak2018candle}, and deliver more accurate and faster results. At present, ML is being implemented in a wide variety of industries and various domains including self-driving cars \cite{bojarski2016end}, cyber-fraud detection \cite{kumar2006managing}, online recommendation systems \cite{wei2017collaborative}, pattern and image recognition \cite{simonyan2014very} and many more examples. This led researchers to investigate ML as a potential solution to automate and handle enormous, complex issues in scientific domains.

ML techniques typically aim at enabling the machine to automatically learn and predict through the process of training and evaluating with respect to some objectives. ML algorithm learns from the present data in the training phase and predicts the task based on the future data. A larger volume of data in the training phase ensures higher accuracy of the predictive models. On the one hand, ML adds an intelligent learning capability for uncovering the underlying patterns in raw data. On the other hand, the scientific workflow deals with large-scale datasets and high computations that generate a massive amount of results. Putting these together, a great potential for scientific workflow is to leverage ML as an alternative to manually handle jobs by human interaction. However, to support distributed and heterogeneous datasets (e.g., data with heterogeneous features in bioinformatics), ML methods should be carefully selected, designed and optimized to operate properly in scientific execution. In this paper, we categorize these optimizations into the creation, execution, and output of a scientific workflow. Exploring the literature, these optimizations take various forms that can be roughly categorized into optimizations that improve the experience of users, improve the models used, tune workflow parameters, ease data handling, guard against failures, improve I/O performance, and system performance as a whole. Therefore, it is important to understand the role of each part for applying suitable ML algorithms.

\section{\textbf{Creation of a workflow}}
Creation of a workflow can be characterized in multiple dimensions: preparing input data, models and parameters, and composition of a workflow. If required, raw data from multiple sources are prepared and transformed into the right format for subsequent execution. These datasets can be unstructured, incomplete, and inconsistent, so need to be indexed and annotated for further accesses. In the second part, parameter tuning and computational steering are discussed and current ML solutions are highlighted. In the last part of this section, we discuss the importance of workflow composition, different methods to create them, and ML solutions to ease this process.
\subsection{\textbf{Prepare input data}}
\begin{figure*}
\centering
    \includegraphics[width=0.9\textwidth]{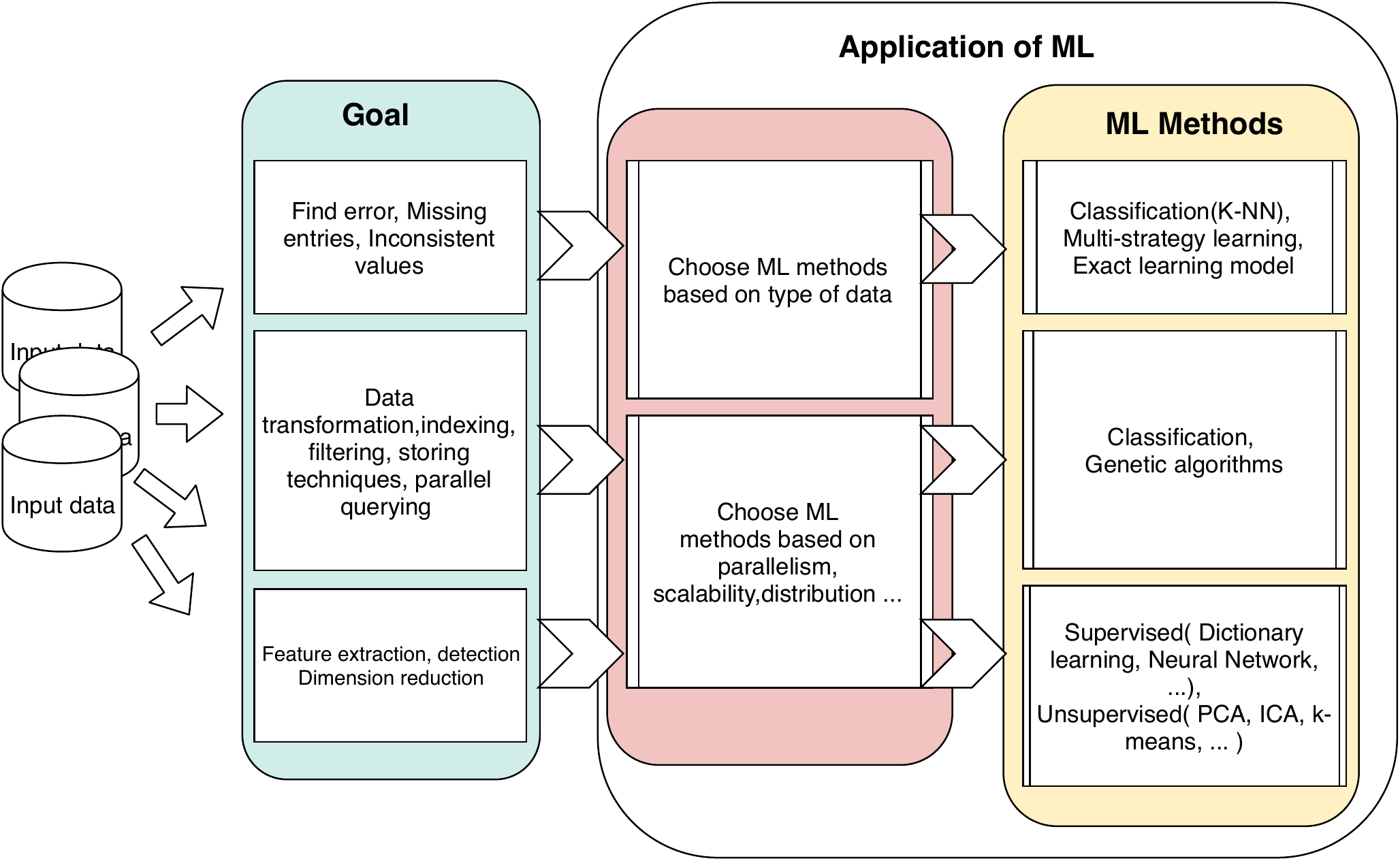}
    \caption{\textbf{Identified challenges related to data and corresponding ML-based solutions.}
    Three main tasks related to preparing data for simulation/analysis/visualization and proposed ML methods in the literature.
    Choosing the right ML method can be based on type of data or characteristics of method itself to satisfy some needs including scalability, parallelism, and so on.}
    \label{fig:Prepare}
\end{figure*}

Scientific workflows deal with a large volume of data and in the case of modern scientific applications running on HPC, they can even produce up to peta-scale data \cite{zheng2010predata}. To gain insight from this vast amount of datasets, exploring the smaller region of interest, complex data analysis, or visualization have been widely used \cite{ayachit2015paraview}. However, sheer volume of data prevents fast query or storing. Therefore, parallel file systems \cite{welch2008scalable} and I/O middleware \cite{polte2008fast}, parallel indexing and querying systems \cite{chou2011parallel}, were proposed to provide efficient access to the scientific workloads. Datasets prior to (or during) the execution of workflows, are prepared, characterized \cite{rubel2008high}, organized, indexed \cite{karpathiotakis2014adaptive}, sorted, or filtered to facilitate further data analysis or visualization \cite{zheng2010predata}. Here ML can be applied to recognize or predict the location of essential information, reduce the precision of retrieved data for exploratory process when it is not needed (ex. combustion simulation \cite{wu2009fastbit}), extract the most important features of data, classify the data that gain insight for visualization and data analytics, or find the correlated data records and search operations \cite{wu2009fastbit}. Also, in-situ middlewares \cite{docan2012dataspaces}, initially targeting I/O, have been used to accelerate preparation of data for the visualization, and data analytic \cite{dreher2016bredala}. 

Analyzing domain-specific data may require investigating multiple files and, in the case of simulation results, thousands of raw data files. As is shown in Figure \ref{fig:Prepare}, raw data often needs to be reformatted for related analysis since it should be in the format of particular analysis programs. Raw data can contain semantic errors, missing entries, inconsistent values, unresolved duplicates, or inconsistent formatting, thus it needs to be sanitized and transformed into usable data prior to analysis \cite{kandel2011research}. This process can be done manually or through scripts in programming languages such as Python, Perl, and R. Reformatting raw data can lead to new insights about the characteristics of data and required assumptions appropriate for the subsequent simulation/analysis. In some domains, it is required to integrate and synthesize complex data to use it in scientific workflows \cite{bowers2004ontology}. Heterogeneous data transformation and integration systems across multiple data models become more difficult as the number of resources, analytical tools, and computational services increases \cite{klien2007rule}. The main challenge here is providing support for the heterogeneous and distributed data stored in different formats, dealing with high dimensional data where scalability, and efficiency of methods play an important role. For example, data obtained in bioinformatics, scientific instruments, astronomical, and sensors show various data formats and representations. To facilitate the handling of various data formats and relevant data elements, feature selection (e.g., genetic algorithm) and classification algorithms (e.g., decision tree, support vector machine, neural network, logistic regression and k-nearest neighbor) can be considered as good candidates. The process of feature selection for large-scale datasets can take a large amount of effort if based upon human expertise. Therefore, an automation approach at scale can optimize this long process and reduce the need of human interaction. For example, ensemble feature selection is able to extract important features in a distributed environment. While ML algorithms can greatly reduce the overhead of handling such large datasets, a lack of domain knowledge can lead to using the wrong algorithm and consequently obtaining unstable results. ML models can be constructed based on the characteristics of different data models and domains. Figure \ref{fig:Cycle} summarizes the different applications of ML for achieving correction, transformation, and extraction of important features of input data.

\subsection{\textbf{Models and Parameters}}
In scientific workflows, users who possess domain expertise need to consider and modify several configuration parameters (data elements, datasets, attribute values, data transformation, tolerance, and error thresholds) for the execution of a computational problem or simulation that they are trying to solve or execute \cite{souza2019keeping}. Due to the difficulty of determining the right values for these parameters before the execution, dynamic workflows have been used to allow for fine tuning ongoing experiments \cite{danani2015computational}. This allows users to dynamically change specific values for the parameters of the workflows while running on HPC systems. This can be observed in Figure \ref{fig:Cycle}. Online data analysis and online adaptation help data-driven decision \cite{nguyen2015workways}, and are supported by monitoring tools \cite{mattoso2015dynamic}. Given this support, users can modify the configuration, checkpoints, datasets, and values of parameters for execution. However, a large number of parameters and the combination of different values can easily complicate the online user steering and confuse the users if these adaptions and values are not properly monitored or registered \cite{souza2019keeping}. Keeping track of each adaption (changes along with values and track of these events) as well as their order can reveal their effects on dataflow as well as workflow performance such as run-time and resource consumption \cite{souza2018provenance}. This is also beneficial for users to understand what input parameters can significantly impact the results and what is the influence of specific input value for the parameters on the output \cite{bauer2016situ}. The need for an online monitoring system that can relate parameter configuration of the workflow to their effects, can be resolved by adopting ML models into the online monitoring systems. Thus, it is important to have an online mechanism that collects parameters and registers user steering data and correlates the output results with that configuration. As an example, ML models in scientific domains deal with large and high-dimension data, resulting in millions or billions of parameters for training large-scale datasets. This training is complex, time consuming, and demands parallel computations. An obvious question for data-intensive machine learning is how to efficiently manage these parameters and how to excavate complex behaviors hidden in large datasets while executing in a distributed environment. Active learning \cite{cohn1996active} can be used to obtain an optimal subset of training data for such ML models where there exist large amounts of unlabeled data for modeling the problems \cite{cockrell2019nested}. This leads us to the need for parallel and scalable scientific ML, considering various factors including energy saving, accuracy, and computational cost, batch size, and learning rate. There have been some studies focused on combining exascale computing with machine learning to address a range of loosely connected problems in cancer research \cite{wu2019performance} aims at testing hyperparameter optimization techniques for ML models. Moreover, studies have attempted to integrate ML and scientific workflow \cite{jungermann2009information}. A recent study \cite{xin2018accelerating} attempted to shorten the time to obtain deployable scientific machine learning models from scratch. The automation of iterative processes, including building, testing, and refining models, can eliminate the need for manually performing such repetitive and lengthy tasks. Accelerating the lengthy process of choosing the right parameters can drastically reduce the time.

\begin{figure*}[h!]
\centering
    \includegraphics[width=0.9\textwidth]{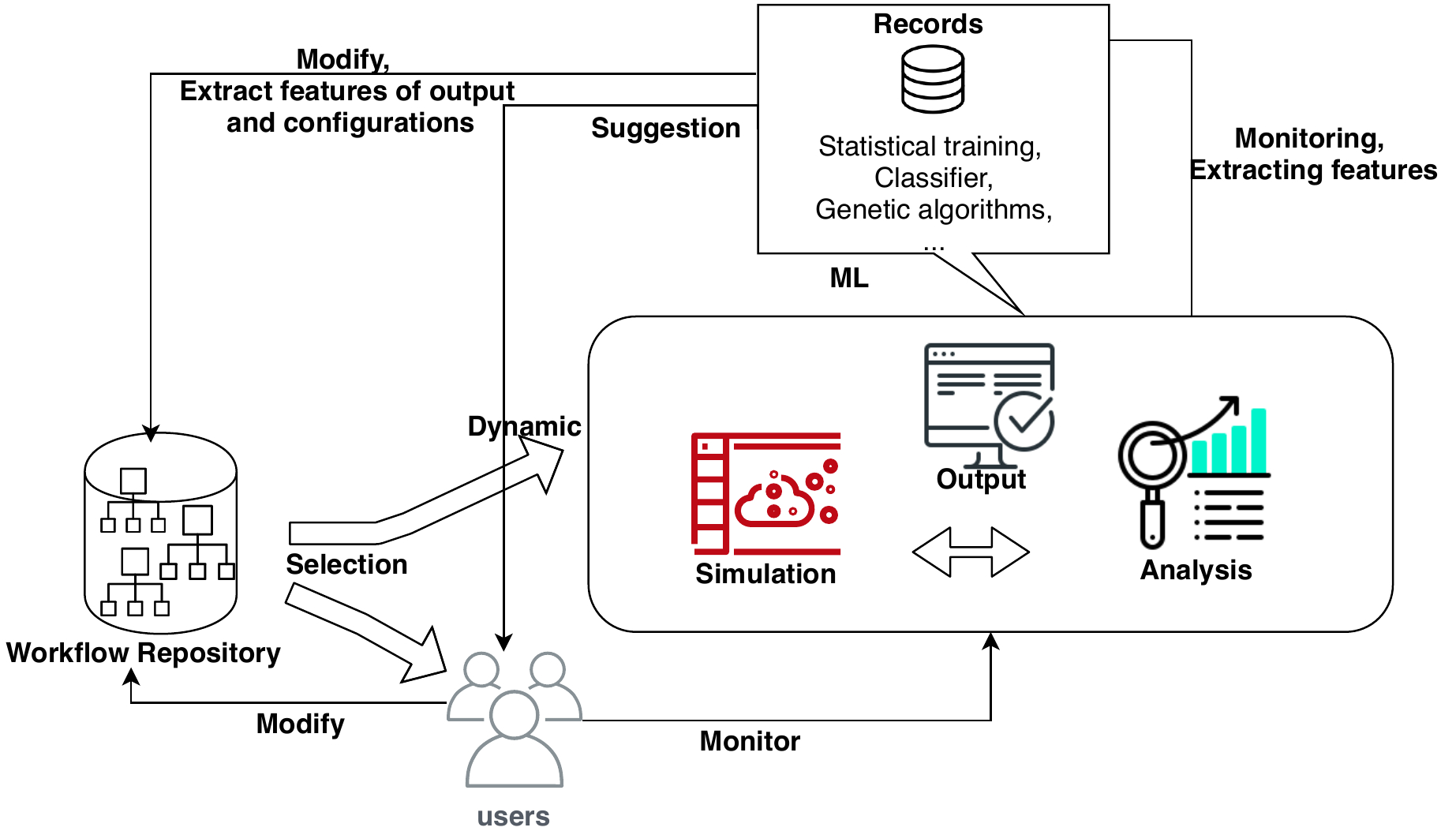}
    \caption{\textbf{The role of ML in the workflows composition system.} Workflow can be selected from repository by user, or dynamically by WMSs. This composition can be modified based on the results obtained from simulation/analysis. Here, ML can extract features from the user's request, compare and map those features against output result, track the history of compositions and provide an optimized composition or a suggestion for future workflow composition.}
\label{fig:Composition}
\end{figure*}

\subsection{\textbf{Workflow Composition}}
Scientific workflows run on large-scale data, consisting of thousands of smaller jobs with data dependencies between each other. Generally, scientific workflows can be represented as Directed Acyclic Graphs (DAGs) \cite{deelman-fgcs-2015} or Directed Cyclic Graphs \cite{lin2011scheduling} (non-DAGs) in workflow models. DAG-based workflows can be scheduled for serial execution or for concurrent execution. Figure \ref{fig:Composition} illustrates the process of composing a workflow. As it is shown in this figure, the process of composing a workflow starts with specifying the data movement across different components and assembled either using a tool that is provided to the user (user-directed), or automatically while it can be chosen from workflow repositories. User-directed composition of a workflow reduces the complexity of the workflow design process and provides the user the ability to reconfigure and change the components of workflows. The execution of large-scale scientific workflows tends to be highly repetitive, both in terms of their iterative nature and in terms of the workload spread across tasks. Users may often desire to use automated workflow composition \cite{gil2011semantic}, through which they can choose and compare the desired configurations among past ones \cite{goble2010myexperiment}, specify different parameters and models \cite{garijo2011new} for creation and execution of a workflow or in an automated or semi-automated process to save time and effort during workflow composition. Users are able to edit these configurations either graphically \cite{taylor2007triana}, or using language modeling \cite{oinn2004taverna}. To specify the right combinations of components that are consistent and complete as a workflow, ML can be a good candidate. The question is how to verify the result of such assisted workflow composition in an automated way and make sure it achieves an optimized choice among all possible configurations. This requires considering input data, domain specific variables, tracking the data movements and corresponding outputs, storing previous execution results and models, and finally the relevant configuration. Researchers need to define intelligent learners that describe and consider all the above parameters and behaviors of existing workflows by applying different input-model pairs and comparing their outputs. However, validating all possible scenarios of input-output is not feasible due to the large space of scientific workflows. ML can be used to give some guides to the selection of the right input and models after considering the result of past choices in an unsupervised manner. In an effort by \cite{piSVM}, the proposed approach eases the limitation of manual job scripts for handling a large number of parameters and finding the right combination of them. In this approach, statistical training is used to find the most important relationship between the features and output variables. 

\section{\textbf{Execution of a workflow}}
\subsection{\textbf{Scientific Workflow Management System}}
\begin{figure*}
\centering
    \includegraphics[width=\textwidth,trim={0 3cm 0 0},clip]{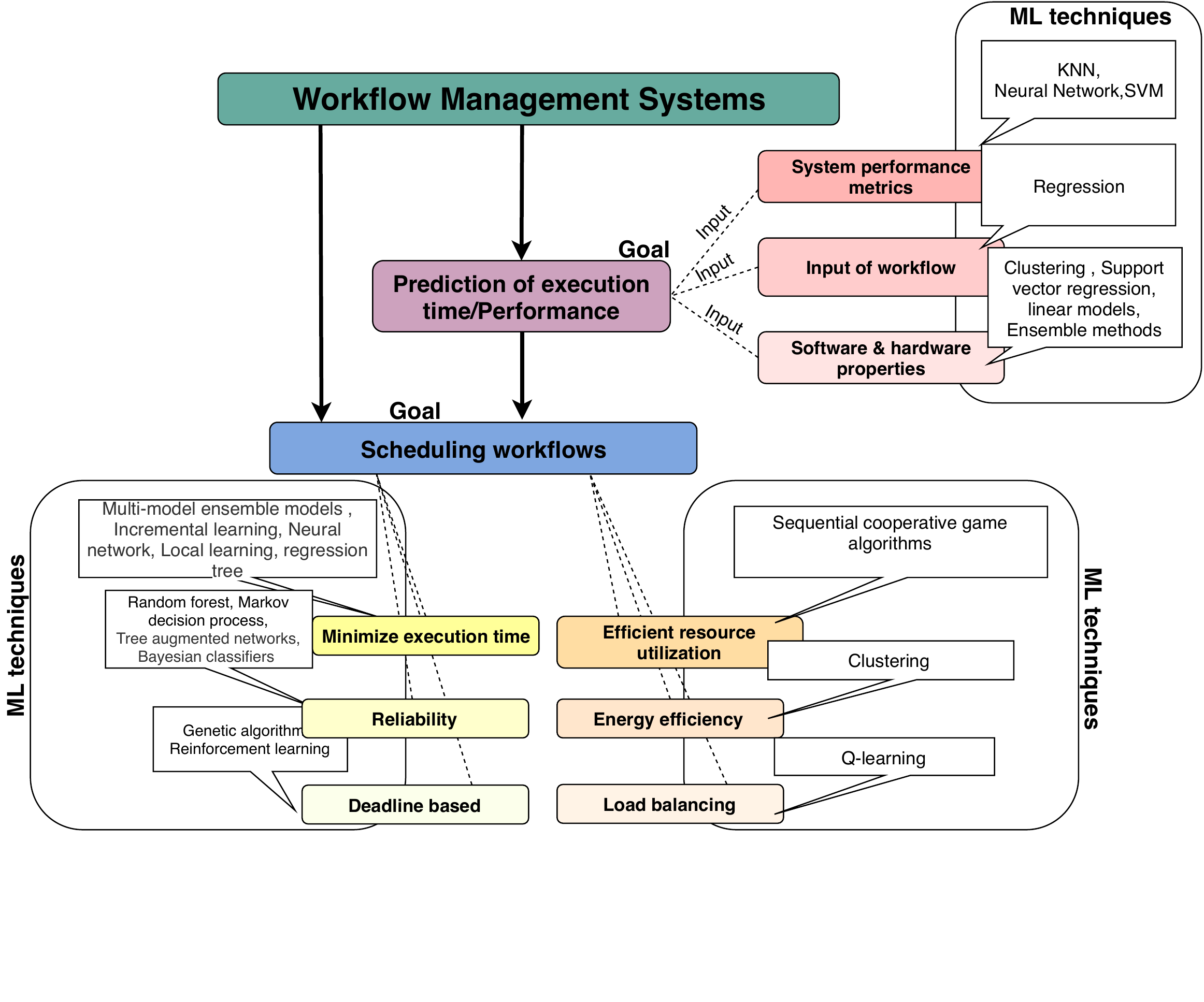}
    \caption{\textbf{Summary of ML techniques for scheduling and predicting runtime.} A WMSs may have multiple objectives for a scheduling policy. For each objective a ML-based scheduling policy is provided. On the bottom of the chart, a summary of goals and the equivalent ML-based solutions are provided. Resource usage and runtime prediction can also be taken into account for designing an optimal scheduling policy. Prediction can be based on characteristics of application and/or underlying system.}
\label{fig:WMS}
\end{figure*}

At each step of a scientific workflow execution, different tasks with various data dependencies are executed. Coordination of these steps and the complex data flow among them are handled by the workflow management system (WMS) \cite{deelman-fgcs-2015}. As a main part of WMS, scheduling of distributed execution on heterogeneous resources aims to minimize the overall execution time while managing the complex problem of experiments over multiple steps \cite{deelman-fgcs-2015}. Scheduling multivariate systems needs the measurement of time for different components such as process I/O, runtime, memory usage, and CPU utilization, or application characteristics, where this information later are used to automatically characterize workflow task requirements, or application runtime \cite{da2015online}. These approaches estimate the resource usage of each task or application's runtime based on their input size, workflows' structure, or characteristics of tasks/application. Their estimation models rely on investigating the profile information obtained from the previous workflow's execution. ML techniques such as ensemble learning, classification trees, and regression trees also have been used to autonomously generate accurate performance models for scientific workflows. Utilizing ML helps to achieve higher accuracy for predicting execution time, memory, and disk consumption for each subtask \cite{hilman2018task} of workflow or application as a whole \cite{juve2013characterizing}. Predicting resource usage can be either at run-time \cite{chen2016uncertainty} or over multiple executions for a single workflow or multiple workflow executions \cite{rimal2016workflow}. The prediction of execution time for each component of a scientific workflow helps to achieve better scheduling policies. However, with the complex dependencies between tasks and data flow at each step (as well as many other factors in distributed systems), it is an open question as to whether it is possible to deliver a scheduling policy that optimizes execution based on all mentioned objectives. In the following, we discuss the two main roles of WMS: scheduling workflow and prediction resource usage. In Figure \ref{fig:WMS}, two main roles of WMS have been highlighted: scheduling and prediction resource usage. In the left side of this figure, various policies for scheduling and their equivalent ML techniques for optimization have been shown and the right side summarizes the techniques that have been used for prediction and ML-based techniques.

\subsubsection{\textbf{Scheduling}}
Scheduling workflows in heterogeneous and distributed environments plays a crucial role in improving the efficiency of resource usage and is known to be NP-complete \cite{ullman1975np}. Additionally, there is no single strategy that is suitable for scheduling distributed systems with different objectives \cite{yu2005taxonomy}. As it shown in the Figure \ref{fig:WMS} exploring scheduling algorithms in the literature suggests different constraints \cite{kalra2015review}. First category of constraints from the user perspective such as: minimizing execution time \cite{barga2008efficient}, deadline based \cite{rodriguez2014deadline}, and reliability \cite{wang2011optimizing}. Second category of constraints from service provider perspective: resource consumption \cite{lin2014hybrid}, load balancing \cite{jianfang2014optimized}, and energy efficiency \cite{zhang2017task}, where some methods supports multiple criteria \cite{li2016scientific}. Most of the approaches are based on heuristic and metaheuristic techniques. It is necessary to present comprehensive algorithms that consider multi-objectives optimization problems with multiple constraints. To provide an optimized solution to the problem of scheduling multi-objective workflows, several models have been proposed using ML algorithms \cite{mao2016resource}. Due to the similarity of mutli-objective optimization problem with multi-agents game theory, scheduling of scientific applications has recently been tackled with ML and Deep Learning methods \cite{xue2017adaptive}. As an example, scheduling multiple large-scale applications while satisfying storage constraints using sequential cooperative game algorithm has presented in \cite{duan2014multi}. This approach formulated the scheduling problem of individual applications as a sequential cooperative game by controlling cost, communication- and storage-aware optimization. Reinforcement learning and Q-learning-based algorithms have been proposed to the problem of scheduling one/ multi workflows in distributed systems by considering the heterogeneity of the nodes, the number of workflows, or the arrangement of tasks in a DAG of dependencies, with the aim of providing better execution time, or minimizing the load balancing \cite{orhean2018new}. Using advanced learning algorithms, such as reinforcement learning, that can independently communicate with the environment, fasten the approximation of an optimal decision and update with the changes, may improve the efficiency of scheduling algorithms in favor of desired objectives. Continuously monitoring the resources and tracking the execution of workflows in distributed systems are now achieve-able through intelligent learner \cite{kintsakis2019reinforcement}. In the light of recent advances in scheduling scientific workflows using ML, there is still further research that needs to be done to establish the role of different ML techniques in distributed environments with single or multiple workflows execution and diverse objectives. 

\subsubsection{\textbf{Predicting Execution Time }}
In scientific applications, properly mapping different workflow components to resources can ensure better performance in a shared environment. Therefore, good scheduling requires an accurate estimation of resource consumption for each task \cite{yu2005taxonomy}. In a distributed environment, estimating the resource consumption of different workflows' components can help the scheduler to better arrange and execute them. In the last decade, there has been a growing interest in predicting the execution time of scientific workflows. The literature on predicting the execution time shows a variety of approaches: system profiles (e.g., system variables and time spent for each state of the system used to construct the prediction model, or the amount of computation per processor and memory hierarchy used to construct a network simulator) \cite{pllana2005performance}, analytical models (e.g. prediction done by solving queuing network models and using historical method, or static performance analysis technique for the external and internal contention on the use of resources by the tasks) \cite{bacigalupo2005investigation}, historical data (e.g. using spatial and temporal composition information useful for performance analysis) \cite{mayer2003iceni}, simulation approaches (e.g using parallel discrete event simulation techniques) \cite{zheng2005simulation}, statistical methods (e.g. predicting the execution time of processors using execution time of small subset of processors along with regression-based approaches to predict parallel program scalability, or time-series models for formulating and predicting the dynamic behavior of complex systems through observations made sequentially through time) \cite{liu2008forecasting}. In recent years, different ML-based scheduling have been proposed to provide accurate estimates of execution time \cite{wu2011performance}. The prediction can be based on system performance attributes \cite{singh2017modular}, or application's input data \cite{pietri2014performance} or combination of both software and hardware properties \cite{monge2015ensemble}. As a simple ML method, regression method has been used in \cite{Miu2012}, where it tries to find a relation between input and tasks' run-time. In this case, selected features of the input (data, hardware characteristics, memory, and etc.) were used as independent variables to predict the performance as a dependent variable \cite{pham2017predicting}. More advance ML algorithms for runtime prediction have been explored such as Regression Trees \cite{da2015online}, neural networks \cite{pham2017predicting}, Long short term memory (LSTM) \cite{hilman2018task}, clustering techniques \cite{pham2017predicting}, Support Vector Machine (SVM) \cite{monge2014ensemble}, or combination of several of these methods \cite{nadeem2019using}. Also, ML has been used to predict run-time of workflows over multiple executions \cite{monge2014ensemble} (using similar features of executions for Ensemble ML technique), or in an offline learning mode (using parameters such as task resource consumption, system configuration and input data). As offline learning methods require the entire dataset to be available for processing in a dynamic environment, where data are continuously generated, new approaches for leveraging ML techniques should be investigated. As an example, an incremental learning approach has been suggested \cite{hilman2018task} to achieve higher accuracy by considering parameters reflecting runtime information such as CPU utilization, memory usage, and I/O activities.

\subsection{\textbf{Scientific simulation and analysis}}
With the flood of data available today, simulation of real-world problems attracts more users than ever before. Simulations on HPC systems produce a large amount of output. Thus, tools are required to manage these large-scale simulations and consequently their results \cite{mattoso2015dynamic}. As HPC simulations are getting larger and more complex, an autonomous scheduler that monitors the resources and jobs at each step and provides real-time decisions for optimizing the resource usage, is required. Additionally, in some domains, users may not know the optimal value for the parameters of the simulation in advance (e.g., the best e-value in bioinformatics), or desire to change the parameters during the execution until the expected result is achieved or to refine the outputs, and lastly if the results is satisfactory then stop the execution. These concerns have been addressed in the literature such as dynamic workflows, user steering, and computational steering \cite{jain2015fireworks}. The parameters for these simulations can obtain various values and reflect different scenarios. Due to the size of data and the complexity of jobs, applying all possible values to the parameters of an experiment is not a feasible solution \cite{danani2015computational}. Therefore, the optimization process for these parametric experiments are considered in various aspects. For example, a subset of values can be selected and explored \cite{silva2018jobpruner}. In this regard, selecting a subset of values and optimizing the parameters can be observed as a mathematical problem and solved by employing ML techniques. Using learning strategies, the scheduler can find the effective parameters and their optimized values to reduce the search space by comparing the past results and identifying unnecessary jobs, thus speeding up the experiments \cite{silva2018jobpruner}. Also, real-time interactive control by which the user can modify the progress of execution, change its parameters and investigate the results through analysis or visualization, has been reflected in the literature \cite{harwood2018real}. Similarly, parameter reconfiguration has been available in workflow management systems \cite{deelman2009workflows}. Users can gain insight about simulation's behavior by exploring different parameters and models \cite{reuillon2013openmole} and representing their effects via visualization \cite{endert2014human}. Therefore there have been many efforts to enable users to interfere \cite{dias2011supporting} or stop the execution or change some parameters, data or the execution \cite{pesic2008constraint} and initiate a new workflow \cite{pradal2015openalea} which necessitate providing some tools that let the users to interact and monitor the execution \cite{vahi2012general}. Also, in-situ processing has been used to accelerate the loop of simulation-analysis by eliminating the need of writing intermediate data back into files for analytic \cite{ayachit2016performance,zhang2016wowmon}. Finally, ML techniques have been utilized to perform CFD simulation in a parallel and online mode \cite{li2019building}. For computationally expensive simulations, ML techniques can prioritize jobs, reduce the search space, predict failure in simulation \cite{jiang2016supervised} and help with prediction in favor of faster simulations.

Developing efficient methods for preparing subsequent data exploration, visualization, or detailed analysis \cite{schadt2010computational}, either manually or automatically \cite{pandey2018event} can be useful to science end users and can gain insights into characterizing data 'in-transit'. Users need to manipulate high-dimensional data \cite{mork2015contemporary} for their peta-scale applications, and the desire to uncover hidden patterns and extract informative features \cite{ferguson2011nonlinear, nouri2021scalable} from these massive datasets for subsequent simulation/analytics \cite{ahmed2007time}. In order to achieve this, staging nodes have been used to reduce the latency of accessing simulations' output for selective data manipulation by offloading output data and exploiting computational power of staging nodes \cite{zheng2010predata}. Data analysis as a main part of scientific workflow, becomes impossible when datasets are large and heterogeneous. ML comes as a solution by proposing clever alternatives to analyzing large volumes of data. By developing fast and efficient algorithms and data-driven models for real-time processing of data, ML is able to produce accurate results and analysis. The more information is fed into the algorithms, the more accurate they become, as the scalability of data and information increases, it allows for much more accurate predictions than ever before. Machine (Deep) learning has been used to cluster scientific data \cite{jiang2019deep}, and dealing with high dimensional data \cite{jiang2019deep}. It is also possible to integrate learning algorithms with workflow management system \cite{nguyen2016integrated} in a distributed mode \cite{wang2014scalable}, construct a machine learning workflow \cite{camillieri2016towards} based on various objectives and input-data. This helps users to easily edit, add, and compare new algorithms to achieve their desired goal.

\subsection{\textbf{I/O and memory optimization}}
Due to the unprecedented pace of scientific discovery, scientific applications are becoming more complex and data-intensive \cite{kitchin2014big}. Simulations generate massive amounts of data \cite{dong2016data}, which must be handled by HPC I/O subsystems. Executing collections of complex processes, which requires transferring large amounts of data from one task to another (as well as intermediate data back and forth between the compute node and the storage system) can introduce a significant I/O bottleneck \cite{dongarra2011international}. As scientific experiments grow to exascale, the performance of these workflows are limited by the I/O constraints imposed by handling large-scale storage systems and complex memory hierarchy \cite{subedi2018stacker}. As a result, data movements between heterogeneous and multi-layers storage systems becomes a main challenge \cite{shoshani2009scientific}. Many I/O middle-ware were introduced to mitigate the difficulty of handling complex memory hierarchy from the users such as ADIOS \cite{lofstead2008flexible}, staging of intermediate data using memory hierarchy, DataSpaces \cite{docan2012dataspaces}, using shared burst buffer, and local burst buffer. Slow reads and writes can drastically decrease the performance of high-speed and low-latency storage systems \cite{caulfield2010understanding}, as in scientific workflows data are generated by one component (e.g. simulation) and will likely be consumed by other components (either simulation or analysis) \cite{li2014tachyon}. Data prefetching as a solution has been used to overcome this, by predicting ahead of time spatial and local data pattern \cite{qin2019towards}. Data prefetcher needs to accurately predict at which time, which data will be requested to increase the performance of the system (reduce the gap between computation node and I/O) \cite{chang1999automatic}. Different strategies have been employed to find optimal data placement \cite{cheng2019optimizing}, manage data movement \cite{dong2016data}, reduce data migration within and across different storage layers \cite{wang2017sideio}. Another solution is to use in-situ techniques for compute intensive workflows \cite{bennett2012combining}, result in reduced amount of data migrated between simulation and analytic. Considering storage hierarchy in data placement can prevent common problems such as load imbalanced and resource contentions \cite{kakoulli2017octopusfs}. ML can be used to automatically identify an optimized decision for migrating data, or tasks to minimize their movement overhead across the deep memory hierarchy and the network \cite{subedi2018stacker,subedi2019leveraging}. As a result, the characteristics of storage level and workflows can be given to the intelligent entity as an input, and it will investigate the output of the defined decision in favor of I/O efficiency.

\subsection{\textbf{Failures and anomalies}}
Distributed environments bring this opportunity for large-scale scientific applications to be executed when they generally require massive computational and data resources. Although executing scientific workflows at a large scale is feasible because of heterogeneous and complex distributed resources, they bring up new challenges \cite{snir2014addressing}. Considering the growth in the size and complexity of such resources, failures in the system are inevitable. Consequently, detecting these failures in real time or even predicting them in advance opens up new research directions. It is essential to recognize the repetitive failures by considering the past executions and prevent them in the future ones. Similarly, anomaly is a growing concern that can lead to failure or unexpected behavior such as delayed or lengthy performance and can considerably reduce the efficiency. Anomaly detection is trending to be complicated as a result of having different factors in hardware, software, and configuration. Monitoring such complicated systems is a key role in early detection of anomalies and can help to find the cause and take actions to ease the effects of these unusual behaviors. For time-aware scientific applications \cite{plale2006casa} obtaining correct computational results requires the data to be available and updated in deadline-based manner. Dealing with these challenges throughout scientific workflow execution in a distributed environment needs an intelligent monitoring system that can track changes and identify the failure points in the system in an automated and online mode \cite{gamell2014exploring}. As a part of an effective workflow scheduler, a fault tolerance mechanism should be provided and guaranteed to recognize the possible failures in the system and output results as early as they occur and take actions accordingly for the recovery. Beyond the traditional techniques to handle failures such as create checkpoint \cite{cao2014transparent}, migrate computation closer to data resources \cite{bala2015autonomic}, and automatically initiate another workflow, the current complexity and sheer size of data necessitate the existence of intelligent learners to predict and recover from a failure. ML can help with tracking the changes in resources, availability of data, predicting the pattern of tasks, vs failures, and many other factors to ensure a dynamic fault tolerance mechanism. Identifying and tracking system metrics can leverage ML techniques to detect the cause of failures, and use this information to predict future failures in real time \cite{gamell2014exploring}. In a recent effort by \cite{gaikwad2016anomaly} an auto-regression approach based on statistical methods for online monitoring time-series data of scientific workflows is developed. In this work, they presented a framework for detecting performance anomalies in scientific workflows to improve the reliability of execution, in which anomaly triggers are generated after a certain threshold is exceeded by the error from the predicted auto-regression model. Anomaly detection in streaming data was proposed by \cite{rodriguez2018detecting} in which Hierarchical Temporal Memory (HTM) networks are used as continuous learning systems that learn incrementally in an unsupervised manner. In this approach, HTM model is used for each metric in the system to detect CPU or I/O anomalies in real-time execution. An online parallel ML algorithm for anomaly detection in computational fluid dynamics (CFD) and turbulence analysis introduced by \cite{li2017real}. This study integrates simulation applications with various analysis applications with the use of DataSpaces \cite{docan2012dataspaces} as a flexible interaction and coordination in distributed and virtual shared space. Task failure prediction is considered by \cite{bala2015intelligent} in order to facilitate proactive fault tolerance for scientific workflow applications. Their goal was to design a ML based prediction model that can handle resource and task failures of scientific workflow execution. The proposed model is distributed in two modules where the first module utilizes ML method for prediction of task failure using evaluated performance metrics, and the second module locates the actual failure after execution of the scientific workflow. The prediction models have implemented through Artificial Neural Network (ANN), Naive Bayes, Random Forest, Logistic Regression (LR) and have shown that the maximum accuracy can be obtained by naive Bayes classification method. Another study \cite{padmakumari2019development} on prediction task failure in workflow execution, proposed an intelligent prediction using ML. In this work, the proposed prediction is combined with an ensemble prediction method to improve the accuracy of failure prediction. Again, Naive Byes are considered as the default classifier along with other classifiers including decision tree, logistic regression, multi-layer perceptron, random forest, and bagging techniques. Intelligent predictive mechanism that forecasts the final status of a dynamic job as a success or failure in a particle physics experiment was proposed in \cite{singh2018deep}. The developed framework uses a Deep Learning method known as Long Short Term Memory Neural Network, which performs prediction in real time in the context of scientific workflows and aims to improve the capability of designing fault tolerance systems. In a different approach aimed at finding the best fault tolerance methods to use in scientific workflow management systems, the authors in \cite{guedes2019provenance} proposed a machine learning-based recommender of fault tolerance techniques. Several classification algorithms including decision trees, random forest, SVM, CN2, and naive bayes have examined to produce rules that define a suitable decision module, which result in proactively avoiding failures or detecting them at an early stage. 

\section{\textbf{Output of a workflow}}
\textbf{Provenance} 
Reproducibility of the result, tracking changes of intermediate data, and enabling query information help the users to understand and share knowledge with collaborators and researchers. However, the large amount of data produced by even a simple scientific workflow execution can lead to poor functionality for the users. Providing provenance information that potentially helps the users and encourages them to engage to re-process and refining the execution is a necessity. As a new approach to investigate, ML can produce explainable and reproducible results while reducing the size of information by filtering the informative data and intermediate results vs non-informative ones. ML can highlight useful data and hidden patterns in an effective approach while improving the users' experience. At the same time, scientific users typically modify the models and parameters in a trial-and-error to obtain and explore different scenarios and results. ML as a candidate can extract similar features out of different executions and save time in this lengthy process. This approach has been examined in \cite{silva2018jobpruner} by analyzing previous executions to provide and prioritize jobs for on-going experiments. Using ML for tuning models and parameters of scientific workflows, it can reduce a significant amount of time and human expertise as it is capable of extracting knowledge from large-scale data as well as their underlying patterns. 
 
Among the different aspects of improving the performance of scientific workflow, provenance is increasingly becoming a vital factor in various applications. Provenance is attracting interest due to the importance of reproducing and re-using experimental results of scientific workflows. Taking this into account, the authors in \cite{simmhan2011using} proposed an approach for data quality scoring based on ML technique to construct a quality function using provenance metadata. Accordingly, the presented model and architecture for data quality scoring employs ML as a classifier to effectively determine the data quality score without a complete metadata record. The classifier function in this study is a set of algorithms such as decision tree, MP3, and KStar, which is constructed using the WEKA engine \cite{hall2009weka}.

\section{\textbf{In-Situ workflows}}
Data generated by simulations are traditionally stored in permanent storage for subsequent post-hoc visualization or analysis. However, with HPC enabling exascale computing, simulations now can generate large volumes of data that cannot be stored or exported to the storage system. The gap between computation capability vs I/O performance and capacity is also increasing \cite{kress2016visualization,ziegeler2015situ}. Therefore, the volume of produced data from simulations exceeds the limitations of memory and I/O and at the same time complicates offline analysis and visualization. The large scale of data combined with the continuously increasing complexity of data analysis and visualization, has changed the traditional procedures in favor of processing the analytic/visualization operating in situ \cite{PDAC11,klasky2011situ}. Concurrent approaches based on in-situ have been used to help the data movement from simulation to data analysis pipeline \cite{bennett2012combining,docan2012dataspaces}. In-situ operations mitigate the effect of storing and exploring large volumes of data by enabling analytics that run simultaneously as the data producer on the same nodes. This can help especially with making a real-time decision. In the absence of these capabilities, offline analysis becomes infeasible at this scale. In-situ analysis eliminates the need to transfer entire intermediate datasets to permanent storage, thus it saves storage and I/O. Running analysis/visualization simultaneously with simulation is beneficial when the results of the analysis can give early insights and discoveries about the data and simulations, and is useful for parameter tuning and modifying the models of simulation in flight. The challenges of storing generated data and I/O difficulties for transferring this data from simulation to storage or visualization while operating in-situ have been addressed by presenting different reduction techniques (e.g. compression, multi-resolution output) \cite{nouanesengsy2014adr,li2018data}. However, choosing the right reduction techniques depends on various factors including processing time, loss ratio, feature detection, and finding the most important data to save. To assist with this process, ML can be utilized for classifying different techniques based on their characteristics and connections with the domain feature to provide a better decision. Using ML, one can choose which techniques are more suitable for a specific use case with multiple constraints. For example, if interesting events happen in the simulation based on spatial and temporal locality, ML is able to detect and track them based on given importance feature metrics, then it can save the data accordingly \cite{ling2017using,shead2017embedding}. Furthermore, feature extraction of data has been used for selecting key time steps (based on similarities between time steps) with the goal of storing less data, and achieved by applying deep learning methods in Computational Fluid Dynamics (CFD) \cite{liu2019key}. Also, with the popularity of applying machine/deep learning, new in-situ compression techniques can be explored \cite{peterson2019merlin}. Collectively, ML techniques might be considered for deciding which reduction techniques is suitable based on various domains, use-cases, objectives, and available I/O and memory (e.g. in CFD, deep learning can be a used for selecting time steps \cite{liu2019key}). ML can be part of decision-making for deciding what data might be most informative (in a domain-specific context) for analysis/visualization and then try to compress and preserve the most informative data fields \cite{liu2019novel}. 

The sporadic nature of the analytics has been observed as an opportunity to utilize the idle resources of simulations \cite{zheng2013goldrush}. In particular, by tracking the simulation patterns and identifying the connections between those patterns and the analytics, a prediction model, for estimating the idle time of simulations'/analytics resources, we can optimize the resource utilization. Furthermore, the resource usage of analysis/visualization tasks may vary during the workflow lifetime in some domains \cite{sisneros2016tuned}, which requires an optimal scheduler to adjust its decision according to the instantaneous resource usage. An efficient ML algorithm could substantially improve the decision-making process of a scheduling policy to further approach the maximum resource utilization in the system. Most of the existing techniques do not support dynamic attach/detach of analytics code to/from the simulation, neither they change the node assignment. A few ML-based solutions have emerged, which consider resource usage optimization for in-situ workflows by monitoring data usage, data locality, and memory hierarchies \cite{subedi2018stacker,subedi2019leveraging}. To ensure optimized resource usage, ML can help predicting the type of data transformation for future data analytics or visualization. Therefore, ML should understand the design space of in-situ workflows and decide accordingly. For example, the performance model of algorithms for analysis and visualization can be provided by ML through their characteristics \cite{larsen2016performance}. This is useful to estimate the runtime for each task and decide to process them in-situ or postpone for post-hoc.

\section{\textbf{Summary}}
We categorize the scientific workflow execution into three major stages (see Fig.\ref{fig:Cycle}); steps prior to execution, execution, and result processing. In the first stage, ML can be primarily employed to recognize/predict prominent features of the data, as well as eliminating the unnecessary precision of the data, or classifying the data for subsequent tasks. ML techniques should be flexible enough to work in heterogeneous environments, to deal with distributed data resources and to distinguish various domain-specific requirements. In addition, the effects of parameter configuration on the results can be obtained by adopting ML model in the online monitoring system. This requires a mechanism that collects this information, extracts the possible patterns, and inputs the user. ML techniques can also infer useful information from the historical configurations and use them to optimally assist the process of composing a new workflow in an automated/ semi-automated manner. In the second stage, there has been a large effort for leveraging ML in the execution of a scientific workflow by recognizing the possible patterns of execution, I/O behavior, and resource consumption that can be later used for managing different components of the system/workflow. The best applicable ML technique is an open problem for future studies that can be applied to online/offline learning over single/multiple executions or workflows and the scalability problem. Although ML has been used to tackle the multi-objective scheduling problem, it is of critical importance to design techniques with measurable performance, to further facilitate the comparison between the available techniques. For better scheduling, the resource consumption can be predicted, for which ML can be utilized with the information obtained from the monitoring system and/or the characteristics of tasks. Furthermore, the simulation and analysis/visualization expose several challenges that can also be dealt with using ML techniques. For instance, a simulation may run several times until it generates a desired result. Difficulties arise, however, when experiments have many parameters to tune. To optimize the process of finding the desired combination of values for these parameters, ML techniques can be used. The existing work reveals many different approaches, e.g., choosing a subset of values, interactive modification of parameters in real time, and extracting the key elements of changing patterns. Moreover, data analysis with heterogeneous and large datasets generated by simulation can be cumbersome. Developing efficient methods for handling and preparing large datasets may leverage ML techniques. For example, ML techniques can classify the data, reduce the dimensionality of the data, or compress data to reduce the transfer time between simulation and analysis. For reducing the time of data movement, ML techniques have been considered, where memory hierarchies and their characteristics define an optimized decision for data placement and migration across the system. Besides, detecting anomalies and failures in the second stage, discussed in many studies and reflected in various approaches, for instance, by tracking the changes that lead to the failure. In Particular, the pattern of failures in the previous runs can be detected upon the occurrence in the execution using the available ML techniques. Furthermore, task or resource failures can be detected as well. And finally, in the third stage in Fig.\ref{fig:Cycle}, reproducibility of the results and sharing with collaborators have to be considered. However, the large volume of results makes the reproduction and sharing time-costly. To improve that, the key elements of the input/output should be extracted, which can be achieved by filtering out the useful information, or by discovering the correlation of input parameters and output features. In other words, uncovering the hidden relation of input, output, and execution can be investigated by ML techniques through multiple executions.

\section{\textbf{Conclusion}}
In this survey, we provide a review of machine learning applications in different stages of scientific workflows. We investigate the techniques that have been used in each part of the scientific workflow execution. Furthermore, we complement the discussion with key insights into the ML techniques that can be leveraged for in-situ operations, limitations, and suggestions for improving the performance of in-situ execution. Running scientific workflows, especially in-situ workflows, require the maintenance of ex-scale systems along with configuring a complex graph of dependent tasks running simultaneously on heterogeneous resources. Avoiding failures through runtime, scheduling the tasks to obtain the maximum performance, I/O and memory configurations, and predicting upcoming events for better scheduling are among the most critical aspects of scientific workflow execution. Many works in the literature utilize ML for performance optimization, accurate estimation of resource usage, and failure detection. In addition, ML shows the potential for extracting the pattern of workflows, data access, and hidden models in the execution. It can further be applied to find automatically the optimal configurations with different sorts of objectives. ML can easily come to help for automatic fault detection and making decision accordingly, classifying workflow execution patterns and choose the optimal configuration based on the underlying system or predicting data access pattern and choose the right decisions to mitigate the effect of data movement. Reliability and scalability of ML methods in all the above mentioned can be more investigated in future. Furthermore, in-situ execution can greatly benefit from ML, as data movements are costly. Thus, ML can detect the data access pattern and provide optimal data/task placements. Moreover, by tracking the generated data from simulation, ML is able to predict in advance what analysis or visualization might be interested and prepare the resources and data appropriately.

\bibliographystyle{IEEEtran}
\bibliography{references}
\end{document}